# A Generalized Tunneling Current Formula for Metal/Insulator Heterojunctions under Large Bias and Finite Temperature

Zenghua Cai[1,2,#], Menglin Huang[1,2,#], Peng Zhou[1,*] and Shiyou Chen[1,2,*]

[1]State Key Laboratory of ASIC and System, School of Microelectronics, Fudan University, Shanghai, China

[2]Key Laboratory for Computational Physical Sciences (MOE), Fudan University, Shanghai, China

The Fowler-Nordheim tunneling current formula has been widely used in the design of devices based on metal/insulator (metal/semiconductor) heterojunctions with triangle potential barriers, such as the flash memory. Here we adopt the model that was used to derive the Landauer formula at finite temperature, the nearly-free electron approximation to describe the electronic states in semi-infinite metal electrode and the Wentzel–Kramers–Brillouin (WKB) approximation to describe the transmission coefficient, and derive a tunneling current formula for metal/insulator heterojunctions under large bias and electric field. In contrast to the Fowler-Nordheim formula which is the limit at zero temperature, our formula is generalized to the finite temperature (with the thermal excitation of electrons in metal electrode considered) and the potential barriers beyond triangle ones, which may be used for the design of more complicated heterojunction devices based on the carrier tunneling.

# 1. Introduction

When a large bias voltage is applied on a metal/insulator/metal or metal/insulator/semiconductor heterojunction, the tunneling effect of electron or hole carriers through wide-gap insulators (semiconductors) can be significant if the insulator layer is thin and the bias is large [1-6]. The large bias produces a large electric field and a large band slope in the insulator layer, which reduces the effective distance for the electrons on the Fermi level of the metal electrode tunneling through the barrier at the metal/insulator interface into the conduction band of the insulator layer, as shown in Figure 1(a), so the tunneling current can be large and increases with the bias (electric field).

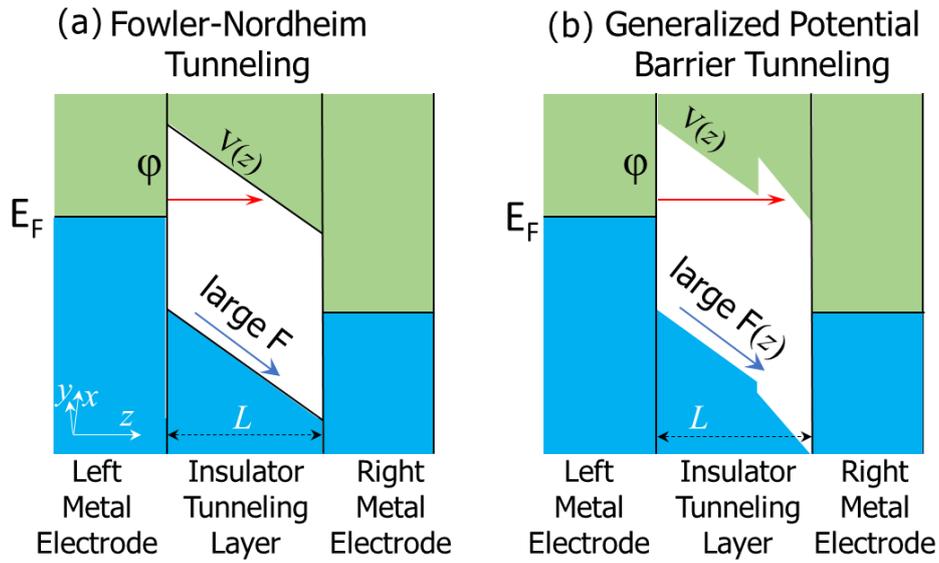

Figure 1. The band diagram of the metal/insulator/metal heterojunctions with (a) the triangle potential barrier which induces the Fowler-Nordheim tunneling effect in the insulator layer, and (b) the potential barrier with the more complicated potential barriers in the insulator layer which is composed of several layers of different insulator materials.

If there is only one layer of insulator material between the two metal electrodes, the potential barrier at the metal/insulator interface has a triangle shape, which has a constant band slope and thus a linear potential barrier function V(x) in the insulator layer. Such kind of tunneling through the triangle potential barrier is called the Fowler-Nordheim (FN) tunneling[1, 4]. A FN tunneling current formula was derived through adopting the WKB approximation to describe the electron transmission function and the nearly-free electron approximation to describe the electrons in the metal electrode [7, 8], in which the current density depends only on the electric field in the insulator layer and barrier height φ at the interface (Equation (2) of [3], (41) of [1] and (12) of [8]). In the derivation, the zero absolute temperature is assumed, *i.e.*, there is no thermal excitation of electrons below the Fermi level to levels above the Fermi level, so the tunneling current is in fact a low-temperature limit (zero temperature) of the current.

If there are two or more sublayers of insulator and semiconductor materials with different band gaps and different dielectric constants in the insulator layer, or the composition of the insulator layer is not uniform, the band in the insulator layer is not flat even under zero bias. In this case, the potential barrier under a certain bias is not the ideal triangle barrier any more, and the potential barrier function V(x) is not linear and becomes more complicated, as shown in Figure 1(b). For such kind of general potential barriers *V(x)*, the FN tunneling current formula is not applicable.

In this work, we adopt the model (as shown by Eq. 2.72 of [9])) that was used to derive the Landauer transport formula at finite temperature, and derive a tunneling current formula that considers the contribution of the thermal excitation of electrons in the metal electrode and is applicable to the more general potential function *V(x)*.

## 2. Derivation of the Tunneling Current Formula

The basic model of our derivation is adopted mainly from Equations 2.70-2.81 of Ref. [9] in which the Landauer transport formula of two-terminal system (the fundamental version of the Landauer-Büttiker formula of the multi-terminal system) was derived and Equations 1-3 and 9-12 in Ref. [8] in which the FN tunneling current formula was derived. If we compare the derivation of Landauer transport formula [9] and FN tunneling current formula [8], we can notice that the physical picture is similar. The major difference is that the zero absolute temperature is assumed in the Fermi-Dirac distribution, the electrons in the metal electrode are explicitly considered as nearly-free electrons and the WKB approximation is used to calculate the transmission function in the derivation of FN tunneling current formula [8], while the finite temperature Fermi-Dirac distribution and an inexplicit transmission function (2.80 and 2.81 of Ref. [9]) are used in the derivation of Landauer transport formula[9].

In the following derivation of our tunneling current formula, we use the nearly free electron approximation to describe the electrons in the semi-infinite metal electrode [9] and WKB approximation to calculate the transmission function explicitly, and meanwhile consider the thermal excitation of electrons in metal electrode through the finite temperature Fermi-Dirac distribution. Now we will show the derivation.

We consider the electron tunneling from the left metal electrode through the insulator layer to the right metal electrode in Figure 1. The tunneling is along the *z* direction and the transverse layer is parallel to the (*x,y*) plane. The Fermi level of the left electrode is $E_F$. There is a potential barrier φ (energy difference between the Fermi level $E_F$ and the conduction band edge of the insulator layer) at the left-metal-electrode/insulator interface, and the potential barrier function V(*z*) (referenced to $E_F$) gives the conduction band edge tilting of the insulator layer. For the FN tunneling with the triangle potential barrier, V(*z*)= φ-F*z* (*z*=0 at the left-metal-electrode/insulator interface and F is the electric field in the insulator layer).

Adopting the nearly-free electron approximation for electrons in the left metal electrode, the energy of the electrons can be written as,

$$E = E(k_x, k_y, k_z) = \frac{\hbar^2}{2m}(k_x^2 + k_y^2 + k_z^2) + V_0 = E_{xy} + E_z + V_0 \quad (1)$$

in which,

$$E_{xy} = E_{xy}(k_x, k_y) = \frac{\hbar^2}{2m}(k_x^2 + k_y^2) = \frac{\hbar^2}{2m}k_{xy}^2 \quad (2)$$

$$E_z = E_z(k_z) = \frac{\hbar^2}{2m} k_z^2 \qquad (3)$$

where $k_x, k_y, k_z$ are the wave vectors (momentum) along *x, y, z* directions respectively, $\hbar$ is the reduced Planck's constant, *m* is electron mass, and $V_0$ is the potential energy reference. $E_{xy}$ is kinetic energy in the transverse (*x,y*) plane, and $k_{xy}$ is defined as the length of the $(k_x, k_y)$ vector. $E_z$ is the kinetic energy along the *z* direction.

Starting from Eq. 2.72 of [9] (the origin can date back to the work of Hans Bethe[10]), the tunneling current $J_{L \to R}$ through the insulator layer from left metal electrode to the right metal electrode can be described by,

$$J_{L \to R} = e \sum_n \int_0^\infty T_{L \to R}(n, k_z) f_L(n, k_z) v_L(n, k_z) \frac{dk_z}{2\pi} \qquad (4)$$

$$= e \int_0^\infty \sum_n T_{L \to R}(n, k_z) f_L(n, k_z) v_L(n, k_z) \frac{dk_z}{2\pi} \qquad (5)$$

where *e* is the electron charge, *n* is the transverse quantum number, $T_{L \to R}(n, k_z)$ is the transmission function of the electron in the *n*-th state with the wave vector $k_z$, $f_L(n, k_z)$ is the Fermi-Dirac distribution which determines the occupation number of electrons on the *n*-th state with the wave vector $k_z$, and $v_L(n, k_z)$ is the group velocity of electron in the *n*-th state with the wave vector $k_z$.

For a given $k_z$, the surface density of electrons with different wave vectors (momentum) in the transverse $(k_x, k_y)$ plane can be calculated as,

$$\int_0^\infty \frac{2}{(2\pi)^2} 2\pi k_{xy} dk_{xy} = \int_0^\infty \frac{2}{(2\pi)^2} 2\pi \frac{m}{\hbar^2} dE_{xy} = \int_0^\infty \frac{m}{\pi \hbar^2} dE_{xy} \qquad (6)$$

In order to calculate the tunneling current density $j_{L \to R}$ along the *z* direction, the summation over the transverse quantum number *n* in the tunneling current formula (5) can be replaced by the surface density of electrons with a given $k_z$, so the *n* summation is changed into the integration over the energy $E_{xy}$,

$$j_{L \to R} = e \int_0^\infty \int_0^\infty T_{L \to R}(E_{xy}, k_z) f_L(E_{xy}, k_z) v_L(E_{xy}, k_z) \frac{m}{\pi \hbar^2} dE_{xy} \frac{dk_z}{2\pi} \qquad (7)$$

According to definition of the group velocity (Equation 2.73 of Ref. [9]),

$$v_L(E_{xy}, k_z) = \frac{1}{\hbar} \frac{\partial E}{\partial k_z} = \frac{1}{\hbar} \frac{\partial E_z}{\partial k_z} \qquad (8)$$

Putting (8) in (7), the integration over $k_z$ is changed into that over $E_z$,

$$j_{L \to R} = e \int_0^\infty \int_0^\infty T_{L \to R}(E_{xy}, E_z) f_L(E_{xy}, E_z) \frac{m}{\pi \hbar^2} dE_{xy} \frac{1}{2\pi \hbar} dE_z \qquad (9)$$

$$= \frac{em}{2\pi^2 \hbar^3} \int_0^\infty \int_0^\infty T_{L \to R}(E_{xy}, E_z) f_L(E_{xy}, E_z) \, dE_z \, dE_{xy} \qquad (10)$$

$$= \frac{em}{2\pi^2 \hbar^3} \int_0^\infty \int_0^\infty T_{L \to R}(E_{xy}, E_z) f_L(E_{xy}, E_z) \, dE_{xy} \, dE_z \qquad (11)$$

According to the Wentzel–Kramers–Brillouin (WKB) approximation, the transmission coefficient $T_{L \to R}(E_{xy}, E_z)$ can be written as (See Equation (1) of Ref. [8] and Equations 3.4.31-36 of Ref. [11]),

$$T_{L \to R}(E_{xy}, E_z) = \left| e^{-\int_0^L \sqrt{\frac{8\pi^2 m^*(V(z) + E_F - E_z)}{h^2}} dz} \right|^2 \qquad (12)$$

$$= \left| e^{-\int_0^L \sqrt{\frac{8\pi^2 m^*(V(z)-(E_z-E_F))}{h^2}}dz} \right|^2 \quad (13)$$

where $V(z)$ is the potential barrier function along the z direction (referenced to the Fermi level $E_F$ of the left metal electrode, as shown in Figure 1(a)), $L$ is the thickness of the insulator layer, $m^*$ is the effective mass. As we can see, $T_{L \to R}(E_{xy}, E_z)$ is independent of $E_{xy}$, so $T_{L \to R}(E_{xy}, E_z)$ can be written as $T_{L \to R}(E_z - E_F)$. Therefore,

$$j_{L \to R} = \frac{em}{2\pi^2 \hbar^3} \int_0^\infty \int_0^\infty f_L(E_{xy}, E_z) \, dE_{xy} \, T_{L \to R}(E_z - E_F) dE_z \quad (14)$$

The Fermi-Dirac distribution of electrons in the left electrode $f_L(E_{xy}, E_z)$ at the temperature $T$ (Note, $k_B$ is the Boltzmann constant and $A = \frac{E_{xy}+E_z-E_F}{k_B T}$),

$$f_L(E_{xy}, E_z) = \frac{1}{e^{\frac{E_{xy}+E_z-E_F}{k_B T}}+1} \quad (15)$$

$$\int_0^\infty f_L(E_{xy}, E_z) \, dE_{xy} = \int_0^\infty \frac{1}{e^{\frac{E_{xy}+E_z-E_F}{k_B T}}+1} dE_{xy} \quad (16)$$

$$= k_B T \int_{\frac{E_z-E_F}{k_B T}}^\infty \frac{1}{e^A + 1} dA$$

$$= k_B T \left[ \ln\left(\frac{e^A}{e^A+1}\right) \right]_{\frac{E_z-E_F}{k_B T}}^\infty$$

$$= k_B T \left[ 0 - \ln\left(\frac{e^{\frac{E_z-E_F}{k_B T}}}{e^{\frac{E_z-E_F}{k_B T}}+1}\right) \right]$$

$$= -k_B T \ln\left(\frac{e^{\frac{E_z-E_F}{k_B T}}}{e^{\frac{E_z-E_F}{k_B T}}+1}\right) \quad (17)$$

Putting (17) in (14),

$$j_{L \to R} = -\frac{em}{2\pi^2 \hbar^3} \int_0^\infty k_B T \ln\left(\frac{e^{\frac{E_z-E_F}{k_B T}}}{e^{\frac{E_z-E_F}{k_B T}}+1}\right) T_{L \to R}(E_z - E_F) dE_z \quad (18)$$

Define $Y = E_z - E_F$ as the incident kinetic energy of electrons along z direction relative to the Fermi level $E_F$, then

$$j_{L \to R} = -\frac{em}{2\pi^2 \hbar^3} \int_{-E_F}^\infty k_B T \ln\left(\frac{e^{\frac{Y}{k_B T}}}{e^{\frac{Y}{k_B T}}+1}\right) T_{L \to R}(Y) dY \quad (19)$$

Following the procedure of Sommerfeld and Bethe used in the derivation of Fowler-Nordheim tunneling current formula, Equation 12 of Ref. [8]), we can extend the low limit of the integral to the negative infinity (The validity of this extension will be discussed in the next paragraph),

$$j_{L\to R} = -\frac{em}{2\pi^2\hbar^3}\int_{-\infty}^{\infty} k_B T \ln\left(\frac{e^{\frac{Y}{k_B T}}}{e^{\frac{Y}{k_B T}}+1}\right) T_{L\to R}(Y) dY \quad (20)$$

$$= -\frac{4\pi em}{h^3}\int_{-\infty}^{\infty} k_B T \ln\left(\frac{e^{\frac{Y}{k_B T}}}{e^{\frac{Y}{k_B T}}+1}\right) T_{L\to R}(Y) dY \quad (21)$$

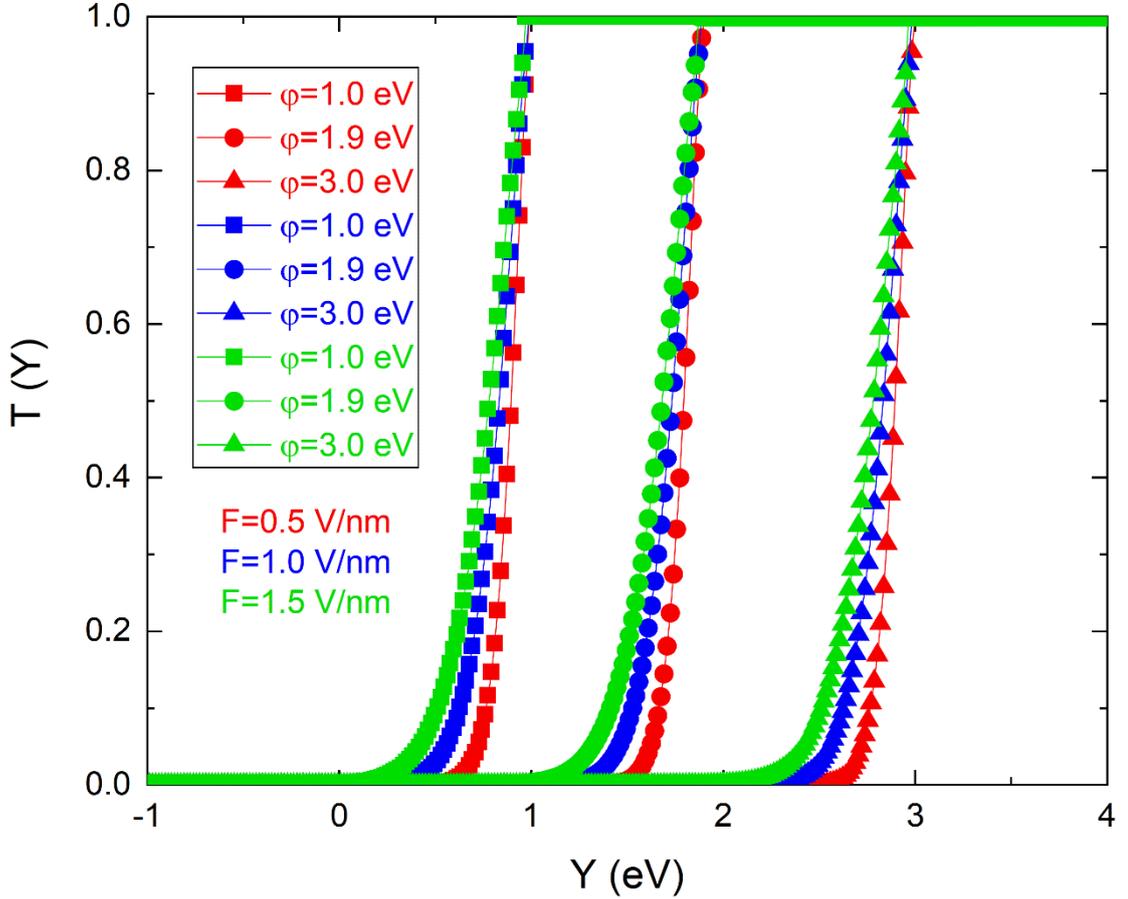

Figure 2. The change of the transmission coefficient $T_{L\to R}(Y)$ as a function of the incident energy $Y$ for the heterojunctions with band diagram in Figure 1(a) which have the barrier φ higher than 1 eV, the electric field smaller than 1.5 V/nm and $m^*$=2.21m.

The validity of the extension is based on the fact that the tunneling possibility decreases very quickly as the incident kinetic energy $Y$ decreases. The transmission coefficient $T_{L\to R}(Y)$ decays quickly to 0 when $Y = E_z - E_F$ decreases to close to 0 and is always 0 for negative Y when the barrier φ is not very small, the electric field $F$ is not very large and the effective mass $m^*$ is not very small in Equation (13). For most of the tunneling devices with wide-gap insulator materials that we studied, this is always valid. Whether this is valid can be tested through a direct calculation of the energy-dependent transmission coefficient function $T_{L\to R}(Y)$, as shown by the test calculations in Figure 2 for the heterojunctions in Figure 1(a) which have the barrier φ higher than 1 eV, the electric field smaller than 1.5 V/nm and $m^*$=2.21m.

Since $Y = E_z - E_F$ is the incident kinetic energy of electrons along the z direction relative to the Fermi level, we can also write it as $E$ (Note, E here is not the E in Equation (1)). The charge of an electron is in fact negative, so the minus sign can be removed for electron tunneling. Then the formula is changed into,

$$j_{L\to R} = \frac{4\pi em}{h^3} \int_{-\infty}^{\infty} k_B T \ln\left(\frac{e^{\frac{E}{k_B T}}}{e^{\frac{E}{k_B T}}+1}\right) T_{L\to R}(E) dE \quad (22)$$

in which,

$$T_{L\to R}(E) = \left|e^{-\int_0^L \sqrt{\frac{8\pi^2 m^*(V(z)-E)}{h^2}}dz}\right|^2 \quad (23)$$

This is a generalized tunneling current formula under the finite temperature $T$ and is valid for any potential function as defined by the potential barrier function $V(z)$.

## 3. Comparison to the Fowler-Nordheim (FN) Tunneling Current Formula

Since this is a generalized tunneling current formula at finite temperature and the physical model used in the derivation is similar to that of the FN tunneling current formula, the results from the generalized formula should be close to the FN tunneling current when the temperature T is low. This can be considered as a test of the correctness of our derivation.

The FN tunneling current formula is (Equation (2) of [3], (41) of [1] and (12) of [8]),

$$j = \frac{q^3 F^2 m}{8\pi h \varphi m^*} \exp\left[\frac{-8\pi\sqrt{2m^*}\varphi^{\frac{3}{2}}}{3heF}\right] \quad (24)$$

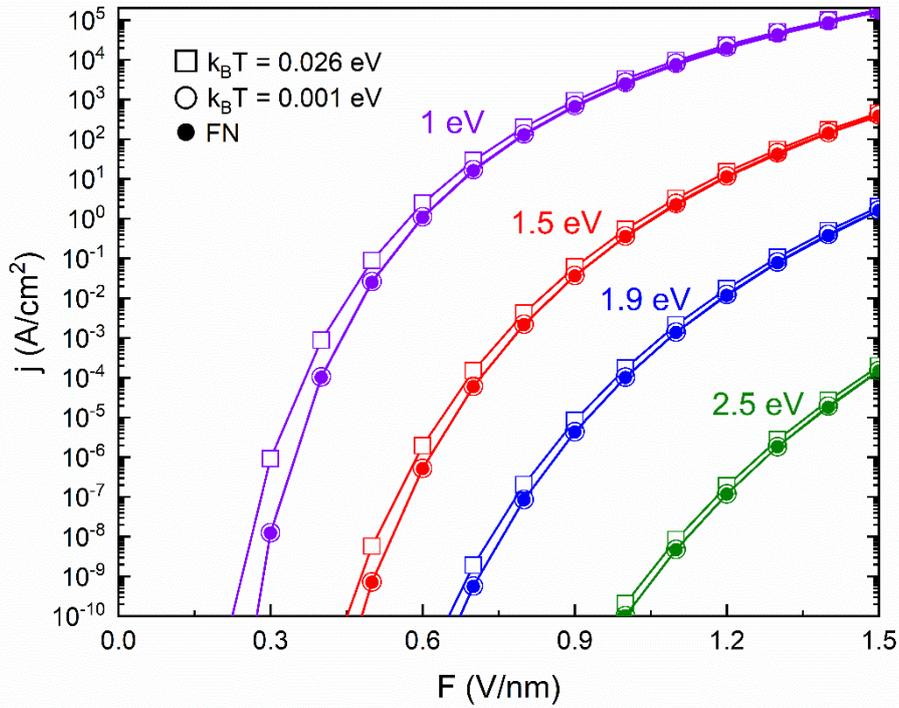

Figure 3. The comparison of the calculated tunneling current density from the FN formula and the present generalized formula under low temperature ($k_B T$=0.001 eV) and room temperature ($k_B T$=0.026 eV). The tunneling current density as a function of the electric field is calculated for the heterojunctions with the band diagram in Figure 1(a) which have four different barriers φ= 1, 1.5, 1.9 and 2.5 eV, and $m^*$=2.21m.

As shown by the results in Figure 3, the calculated tunneling current density for the tunneling through the triangle barrier in Figure 1(a) agrees very well between those from the FN tunneling current formula and those from the present generalized formula under low temperature ($k_B T$=0.001 eV, and T is around 12 K). The points almost overlap for the results from the two formulae, indicating that the two formulae are equivalent at low temperature.

At room temperature ($k_B T$=0.026 eV), the tunneling current from the generalized formula is higher than that from the FN formula, because the generalized formula considers the thermal excitation of the electrons from below the Fermi level to the higher-energy levels which increases the energy of incident electrons in metal electrode relative to the barrier in the insulator layer. When the barrier is large or the electrical field is large, the difference becomes very small. Only when the barrier is small and the electric field is small, the relative difference can be significant. This difference may be important for causing the leakage current in the metal/semiconductor heterojunctions in which the band gap of the semiconductor is not large and thus the barrier is not large, and meanwhile the bias and electric field in the working device are also not large. In such cases, it will be more accurate to adopt the generalized formula at finite temperature.

## 4. Conclusions

Following the model that was used to derive the Landauer formula at finite temperature and adopting the nearly-free electron approximation to describe the electronic states in semi-infinite metal electrode and the Wentzel–Kramers–Brillouin (WKB) approximation to describe the transmission coefficient explicitly, we derived a generalized tunneling current formula for metal/insulator heterojunctions under large bias and with large electric field in the insulator layer. The formula is shown to be equivalent to the Fowler-Nordheim formula which is the limit at zero temperature. The formula is generalized to the finite temperature (with the thermal excitation of electrons in metal electrode considered in the Fermi-Dirac distribution) and the potential barrier functions beyond the triangle ones, *e.g.*, when the insulator layer is composed of several sublayers of insulator and semiconductor materials with different band gaps and different dielectric constants.